%% file: main.tex
\documentclass[conference]{IEEEtran}

\usepackage{authblk}
\usepackage[symbol]{footmisc}
\usepackage{commands}

\begin{document}
\title{Characterizing Developer Use of Automatically Generated Patches}

\author[1]{Jos\'e Pablo Cambronero\textsuperscript{*}}
\author[1]{Jiasi Shen\textsuperscript{*}}
\author[1]{J\"urgen Cito\textsuperscript{*}}
\author[2]{Elena Glassman}
\author[1]{Martin Rinard}
\affil[1]{MIT, Cambridge, MA \authorcr \vspace{-2ex}}
\affil[2]{Harvard University, Cambridge, MA \authorcr \vspace{-2ex}}

\maketitle
\footnotetext[1]{Equal contribution}
\footnotetext{This paper is an extended version of
a short paper, by the same title, that appeared in VL/HCC 2019.
}

\begin{abstract}
\input{abstract}
\end{abstract}

\input{introduction}
\input{methodology}
\input{findings}
\input{implications}
\input{threats}
\input{related}
\input{hci_related_work}
\input{conclusion}

\bibliographystyle{plain}
\bibliography{references}

\end{document}

%% file: abstract.tex
%!TEX root = main.tex
We present a study that characterizes the way developers use
automatically generated patches when fixing software defects.  Our
study tasked two groups of developers with repairing defects
in C programs. Both groups were provided with the defective line
of code. One was also provided with five automatically generated
and validated patches, all of which modified the defective line of
code, and one of which was correct. Contrary to our initial
expectations, the group with access to the generated patches did
not produce more correct patches and did not produce patches in
less time. We characterize the main behaviors
observed in experimental subjects:
a focus on understanding the defect and the relationship
of the patches to the original source code. Based
on this characterization,
we highlight various potentially productive directions for future
developer-centric automatic patch generation systems.

%% file: introduction.tex
\section{Introduction}

Software defects have been a known problem ever since the inception of the field
of software development. Recent research in automatic patch generation has
produced systems that have been shown to be capable of generating correct
patches for a significant fraction of the considered
defects~\cite{long2016genesis,long2016automatic, mechtaev2016angelix,
long2015staged}. Many successful automatic patch generation systems
for real-world applications use a generate-and-validate approach --- the system
generates candidate patches that it then validates against a test suite
containing sample inputs and outputs. While this approach has been shown to
successfully generate correct patches, it has also been shown to generate many
more so-called \emph{plausible} patches that produce correct outputs for all
inputs in the test suite, but incorrect outputs for at least some other
input~\cite{qi2015analysis, long2016analysis}. For this reason, the generated
patches should be examined by a developer before integration into the source
code base.
Despite the need for developer involvement, there has been little research
characterizing the developer workflow and potential productivity improvements of
automatic patch generation in comparison with other alternatives.
For example, there is a long history of defect localization
research~\cite{pearson2017evaluating, abreu2009practical, ali2009evaluating},
which aims to identify defective source code that the developer can then
manually patch.

\textbf{Automatic Patch User Study.}
We present a study that characterizes and compares the developer process in
automatic patch generation and manual patch generation aided by defect
localization. Specifically, we compare two populations of developers: one
provided with the location of the defective line of code and asked to manually
develop a patch, and one provided with five automatically generated patches, all
of which validate against the test suite, all of which modify the defective line
of code, but only one of which is correct. This experiment was designed to model
a scenario, inherent in the use of generate-and-validate automatic patch
generation, in which the developer is given patches that validate but may or may
not be correct.

Our study provides a qualitative analysis of the recurring behaviors observed in
the experimental population.

We
found that subjects in the experimental group exhibit the following
behaviors:\\
{\bf Code and Patch Inspection Times:} On average, experimental group subjects
  spent only \fracinspectingpatches{} inspecting the
  provided patches and \fracinspectingvars{} inspecting
  the occurrence of variables throughout the source code.
  The remainder of the time was spent investigating the buggy source code, tests, and
  supplementary information provided.\\
{\bf Patch Comparison:} Experimental subjects
  compared patches in an ad-hoc manner, placing patch files side-by-side and navigating
  back and forth between different patches.\\
{\bf Small vs. Large Modifications:} On average, experimental group subjects spent only \fracdeletepatches{} on patches that made larger modifications to existing source code, such as removing a branch entirely.
  This reluctance to spend time on patches with large modifications was counterproductive for one of the defects in the study --- the correct patch made a large modification.\\
{\bf No Changes:} 5 of 12 patches submitted by experimental subjects
  featured few or no additional manual changes.

Based on these observations, we formulate directions for future research that
helps developers better understand the defect, the relationship of the candidate
patch to the defect, and overall improve the development process when using
patch generation systems. Examples of such directions include:\\
{\bf Variable Instrumentation:}
		Our study participants spent a substantial amount of time investigating the role of variables in the generated patches to understand how they relate to the original defective code.
		Future systems could aid this process by augmenting variables in the
		generated patches with information provided by program
		slicing~\cite{programslicing} or dynamic information
		flow~\cite{sidiroglou2015automatic, sidiroglou2015targeted}.\\
{\bf Successful Patch Characteristics:}
Previous work has shown that machine learning can be used to successfully rank
correct patches~\cite{long2016automatic} by using code characteristics of the
patch.
Providing such information to developers as they inspect patches may help them
more quickly distinguish correct patches from incorrect patches.\\
{\bf Trace and Influence Summaries:}
     Information about how the patch affects program execution characteristics
     may make the potential impact of the patch clearer. This information would
     be collected during the runs of the original, unpatched program and during
     the validation runs for candidate patches~\cite{head2017writing}.\\
{\bf Invariants:}
	  Previous systems have inferred invariants that characterize successful
	  executions~\cite{ernst2007daikon,kruger2006clearview}. Providing invariants
	  that involve patch variables may help developers better understand the roles
	  that these variables play in the overall computation.

\noindent We summarize our contributions in the following:\\
\indent\textbf{User Study:} To the best of our knowledge, this paper presents the
    first study that 1) asks developers to produce correct patches for application logic defects
    and 2) provides the experimental group with multiple plausible automatically generated patches.\\
\indent\textbf{Experimental Results and Qualitative Analysis:}
    We conduct a qualitative analysis on developer interactions with
    the patches and programs to identify challenges in the program repair
    process. We characterize the main behaviors observed and use this to derive
    implications for developer-centric patch generation systems.\\
\indent\textbf{Future Directions:} Based on the results of our analysis, we formulate potential directions for future patch generation research.

%% file: methodology.tex
\section{Research Method}
We now present the design choices for our user study.

\subsection{Study Participants}

We recruited a total of 12 developers who had at least one year of
experience programming in C or C++ in the last five years, and at
least four years of programming experience overall. These subjects
had development experience in a combination of academic and industrial
settings.
The subject population was drawn from the doctoral Computer Science
program at MIT,
focused mainly on systems or
machine learning research.

\Cref{users:experience} shows a breakout of the experience
levels for each subject, along with group and study allocations.

\begin{table}[h]
\centering
\include{figures/user_c_cpp_experience}
\caption{
Study participants had at least one year experience in C or C++, and most had 2 or more years.
Our study tasks only required experience with basic C concepts.
}
\label{users:experience}
\end{table}

We asked participants to fill out a survey evaluating their
understanding of 42 key concepts in C, which
we based on the GNU C manual~\cite{loosemore1993gnu}. Each question could be
answered, in increasing order of experience, as \textit{``Unfamiliar, would
not understand if you see it in code''}, \textit{``Somewhat unfamiliar, cannot write
it but can read and understand it in code''}, \textit{``Familiar, can write it if presented
with similar code that uses it''}, and \textit{``Expert, confident about writing it independently''}.

We identified two types of participants: ``expert'' and ``non-expert''
based on the number of concepts they answered with \textit{``Expert,
confident about writing it independently''}. We set a threshold of
25/42 questions for ``expert'' status. 8 of the 12 participants
were categorized as experts by our metric and the remaining 4 as
non-experts. The answer threshold was set based on the participants'
response distribution.

\Cref{table:experts} provides the number of expert
answers given by each subject.

\begin{table}
  \input{figures/expert_answers}
  \caption{%
  We classified subjects with less than 25/42 expert answers on a C concept survey
  as ``non-experts''. We balanced
  these individuals across both control and experimental groups.
  }
  \label{table:experts}
\end{table}

\begin{figure}
 \includegraphics[scale=0.4]{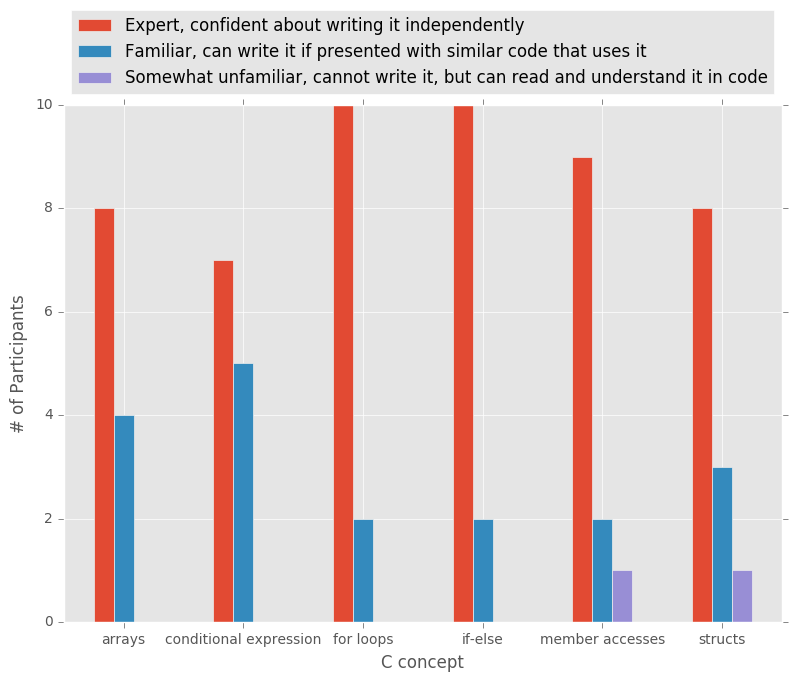}
 \caption{%
 The x-axis presents the C concepts used in the source code
 associated with our user study's tasks. The y-axis presents the
 count of study participants that answered the corresponding survey
 topic with a given answer type.  In all cases, participants
 understood the concepts and most were confident they could write
 such code independently.
 }
 \label{fig:c-concepts}
\end{figure}

\Cref{fig:c-concepts} shows the distribution of subject responses for C concepts directly used in the buggy source code they were tasked with repairing.
Based on their responses, subjects understood how to read all
concepts necessary and in most cases were confident in writing them
independently or when there was an example provided.

We carried out two study variants using this subject population. In the
\emph{long} study, we allotted a total time of 80 minutes per study
participant.  In the \emph{short} study, we allotted a total time
of 45 minutes per study participant. Of the twelve subjects, eight
were assigned to the long study and four to the short study.  The
long study was designed to provide subjects with sufficient time
to solve the bugs without time pressure.  The short study was
designed to observe the effect of patches in a setting where time
pressure was higher for participants.  These time allotments are
consistent with previous studies in active bug repair, which have
allocated a maximum of two hours for five bugs~\cite{paruserstudy}.
We recruited the participants telling them the study ``evaluates
bug fixing tools.''

\subsection{Treatment Groups}

We randomly assigned participants to two groups: \emph{control} and
\emph{experimental}. When doing so, we maintained a balance of
non-expert subjects across treatment groups, based on the answers
subjects provided to the C concept survey. Both control and
experimental treatments each had two non-experts.

\textbf{Experimental:} The experimental group received five validated patches for each of
the bugs. One of the five patches was correct, and four were plausible
but incorrect patches, but subjects were not told that any one patch
was correct. They were given
 instructions and materials that emphasized the potentially
incorrect nature of patches that pass the test suites, which may
occur with current automatically generated patches.  We provided
patches that were generated offline to control for the possible
effect of the tool's interface.

\textbf{Control:} The control group did not receive any generated patches, but did
receive the exact source line that contained the defect. By providing
the exact location of the defect, we ensured that the experimental subjects
and control subjects had the same defect localization information and that this
matched the line modified by the ground truth developer patch.

The sizes of control and experimental groups were balanced in both
long and short studies.

Participants were informed of their time allotment for each bug.
When carrying out the experiment, we divided time slots such that
there was no overlap between study groups in the space we reserved. At
any time, the participants in the room consisted entirely of
experimental or control subjects. Additionally, we informed
participants in the room that not everyone had been provided
with the same experimental setup, so they wouldn't feel pressured
by other subjects' actions (e.g. a subject leaving the room before them).

We provided each subject with a subject id number to avoid
the use of their personal names in study materials. The subject id numbers
start at 2, as we allotted subject id 1 for a pilot subject who
tested the environment, tools, and provided feedback on the instructions.

\subsection{Environment and Tooling}\label{sec:vm}
To simplify the environment setup and encourage reproducibility,
we asked participants to work from a
pre-configured virtual machine.
We standardized the directory
structure for each of the bugs, providing uniform access to relevant
source code and tests to avoid confusion when progressing through
the experiment.  The buggy source file was
linked to a top level file named \path{patch.c} to prevent subjects
from spending time traversing the source directory tree. The bug information
was placed in a top level file named \path{BUG_INFO}. Subjects
were also provided with an IDE alternative (CLion) if they preferred
using that tool over the command line. Subjects were
not allowed to use additional debugging tools to control for levels
of tooling experience.

\subsection{Study Tasks: Repairing Application Logic Errors}\label{sec:bugs}

We based the set of bugs available for our study on benchmarks used
by the literature for automatic patch generation
systems~\cite{long2016automatic,long2015staged,yang2017better}.
To provide participants with reasonable tasks, we focused on bugs that:
were exposed by an accompanying test suite, require application
understanding (to the extent we could effectively explain in accompanying
documentation), and did not require additional tooling (e.g. valgrind) which
could result in skill-set differentials across study cohorts.

Given that our study investigates the potential positive impact of
automatic patch generation, we focused on bugs where our patch
generation system produced at least five validating patches and one
of these validating patches correctly repaired the bug. By having
multiple validating patches, and one correct patch among them, we
sought to exercise the subjects' ability to discern between correct
and incorrect patches.

Our candidate tasks needed to produce validating patches that modified the same
source line location, which in turn had to match the line modified by the
correct developer patch. This allowed us to provide the control group with the
same exact line location information that was provided to experimental subjects
via patches.

This criteria yielded two bugs for the experiment: \libtiffbug{},
an error-checking bug in the popular TIFF library, and \phpbug{},
a bug in PHP's standard library function \verb|substr_compare|.
The relevant portions of the source code for each bug, along with
the appropriate fix, are presented in \Cref{fig:bug:libtiff} and
\Cref{fig:bug:php}, respectively.  The developer patches for
\libtiffbug{} and \phpbug{} are available at ~\cite{libtiffpatch}
and \cite{phppatch}, respectively.

\subsubsection{Task 1: TIFF Image Layout Bug}

To correctly repair \libtiffbug{}, subjects need to understand the
basics of TIFF and its image layouts. They also need to be aware
that the error correcting code executed as a consequence of the buggy
branching condition may not be necessary for all images. We provided
subjects with information in a file titled \verb|BUG_INFO|, which
they were instructed to read prior to starting the timer on each
bug.  The \verb|BUG_INFO| file for this bug is available
online~\cite{onlinebugs}, and contains an overview of TIFF, an
outline of the main data structure relevant for the bug (the
\verb|TIFFDirectory|), along with various input examples that expose
the bug and the expected behavior.

\subsubsection{Task 2: PHP String Comparison Bug}

To correctly repair \phpbug{} subjects must understand the
semantics of the string comparison function \verb|substr_compare|.
In particular, the subjects should understand how comparison works
on different length strings. The \verb|BUG_INFO| file for this bug
is available online~\cite{onlinebugs}. It contains  various input
and output examples for \verb|substr_compare| calls, along with the
function documentation found on the PHP website~\cite{phpdocs}.

\begin{figure}
\begin{lstlisting}[basicstyle=\ttfamily\footnotesize, frame=single, breaklines=True, escapechar=$]
} else if (
  @td->td_nstrips > 1@$\colorbox{pink}{td->td\_nstrips > 2}$
  && td->td_compression == COMPRESSION_NONE
  && td->td_stripbytecount[0] != td->td_stripbytecount[1]
  )
  {
  TIFFWarning(module,
   	"%s: Wrong \"%s\" field, ignoring and calculating from imagelength",
        tif->tif_name,
   		_TIFFFieldWithTag(tif,TIFFTAG_STRIPBYTECOUNTS)->field_name);
   		if (EstimateStripByteCounts(tif, dir, dircount) < 0)
   		    goto bad;
   	}
\end{lstlisting}
\caption{\texttt{libtiff-d13be-ccadf}: The branch condition at line 589 of \texttt{libtiff/tif\_dirread.c} needs to tighten the first
predicate from \lstinline{td->td_nstrips > 1} to \lstinline{td->td_nstrips > 2} to successfully repair the bug, which manifests itself as a custom error when \texttt{libtiff} tries to estimate the strip byte counts for images that don't satisfy the
correct condition.
}
\label{fig:bug:libtiff}
\end{figure}

\begin{figure}
\begin{lstlisting}[basicstyle=\ttfamily\footnotesize, frame=single, breaklines=True]
@if (len > s1_len - offset) {@
  @len = s1_len - offset;@
@}@

cmp_len = (uint)
  (len ? len : MAX(s2_len, (s1_len - offset)));
...
\end{lstlisting}
\caption{\texttt{php-309892-309910}: The branch statement starting at line 5255 in \texttt{ext/standard/string.c} should be removed
to successfully repair the bug, which manifests itself by producing
incorrect output for PHP's \texttt{substr\_compare}.}
\label{fig:bug:php}
\end{figure}

In the \emph{long} study, we allocated 40 minutes for each of these
bugs.  In the \emph{short} study, we allocated 25 minutes for
\libtiffbug{} and 20 minutes for \phpbug{}.

\subsection{Experimental Patches}
Subjects in the experimental group were provided with five plausible
patches per bug. These patches were generated using
Prophet~\cite{long2016automatic}, an automatic patch generation
tool. Other automatic patch generation systems such as
GenProg~\cite{le2012genprog}, SPR~\cite{long2015staged}, and
Angelix~\cite{mechtaev2016angelix}, produce plausible (but incorrect)
patches for these bugs or correct patches that are similar to those
produced by Prophet ~\cite{qi2015analysis, angelixpatches,
yang2017better}.

The patches provided to subjects are available online~\cite{onlinebugs}.
One patch in each set of five was correct and included by design.
The placement order of the correct patch among the five patches
was randomized at design, and every experimental subject was exposed to the
same ordering.

To model practical conditions, we reminded experimental subjects
in both the instructions handout and the tutorial video that patches
produced by the system would all pass the test suites but had no
correctness guarantees.

Each patch provided to experimental subjects had a source code
comment indicating the portion of the code that had been automatically
generated.  \Cref{table:patches} provides details on the modifications
made by each patch. All patches modified the branching conditions
responsible for the bugs.  \Cref{fig:patch:libtiff} and \Cref{fig:patch:php}
present the automatically generated patches that correctly repair
\libtiffbug{} and \phpbug{}, respectively.

\begin{table}[]
\centering
%\resizebox{\columnwidth}{!}{%
\begin{tabular}{@{}lll@{}}
\toprule
Bug     & Patch & Net Effect                                                              \\\midrule
libtiff & 1     & Removed branch by adding false                                         \\
 \textbf{libtiff} & \textbf{2}     & \textbf{Modified predicate to \lstinline|td->td_nstrips > 2|} \\
libtiff & 3     & Removed branch by adding UNSAT predicate                         \\
libtiff & 4     & Added predicate on variable \lstinline|diroutoforderwarning|                \\
libtiff & 5     & Added predicate on variable \lstinline|fix|                                 \\
php     & 1     & Added predicate on variable \lstinline|offset|                              \\
php     & 2     & Added predicate on variable \lstinline|cmp_len|                            \\
\textbf{php}      & \textbf{3}     & \textbf{Removed branch by adding false}                                         \\
php      & 4     & Added predicate on variable \lstinline|cs|                                  \\
php     & 5     & Added predicate on variable \lstinline|ht|                                 \\\bottomrule
\end{tabular}
%}

\caption{%
The net effect of the modifications performed by the patches provided
to subjects in the experimental group for \libtiffbug{} (libtiff)
and \phpbug{} (php). All patches pass the test suite, but only
bolded rows are correct.
}
\label{table:patches}
\end{table}

\begin{figure}
\begin{lstlisting}[breaklines=True, escapechar=$]
else {
    if ((td->td_nstrips > 1 && td->td_compression == 1 && td->td_stripbytecount[0] != td->td_stripbytecount[1]) && $\colorbox{pink}{!((td->td\_nstrips == 2))}$)
{  ... }
\end{lstlisting}
  \caption{%
  The generated patch that correctly repairs \libtiffbug{} adds a predicate equivalent to the developer fix, which modifies \lstinline{td->td_nstrips > 1} to
  \lstinline{td->td_nstrips > 2}, highlighted in pink.
  }
  \label{fig:patch:libtiff}
\end{figure}

\begin{figure}
\begin{lstlisting}[breaklines=True, escapechar=$]
if ((len > s1_len - offset) && $\colorbox{pink}{!(1)}$) {
  len = s1_len - offset;
}

cmp_len = (uint) (len ? len : MAX(s2_len, (s1_len - offset)));
\end{lstlisting}
  \caption{%
  The generated patch that correctly repairs \phpbug{} adds a false conjunct, highlighted in pink, that is equivalent
  to the developer fix, which removes the branch statement.
  }
  \label{fig:patch:php}
\end{figure}

\subsection{Study Instructions}
All subjects received: a PDF document with detailed instructions,
a file detailing relevant information for each bug (including defect line location,
observable errors/reproduction, and API details), and a tutorial video
walking them through the tooling provided. The subjects read both documents
and watched the video before starting the task timers. We also provided them
with a tutorial-related test task prior to starting to familiarize them with the environment.
All documents were tailored based on the group assignment, such that no
additional information was leaked to experimental/control subjects.
We have made these documents available online~\cite{allonlinedocs}.

\subsection{Qualitative Analysis}
The virtual machines used to carry out the experiments
were set up to record all on-screen activities using
\lstinline{avconv}~\cite{ffmpeglibav}.  This totaled nine hours
of on-screen activity across the 12 subjects and two study tasks. These videos
facilitated our qualitative analysis.
In a first pass, two researchers independently collected notes on subject
behavior and identified shared behaviors for use as emergent qualitative
coding~\cite{stemler2001overview}. These behaviors consisted of: searching for a
variable in the source code, reading the \lstinline{BUG_INFO} background file,
and navigating to or editing a source file (in which case, we associated the
file name as the code). These notes and codes were used in a second pass
to qualify frequency of behaviors.

The scripts used were also instrumented to collect time-stamped data in
JSON logs. We used these to provide additional context for our qualitative
analysis.

\subsection{Study Reproducibility}
We provide an extensive replication package to encourage reproducibility of
our study and analysis.
We have made available all the resources used in our study~\cite{allonlinedocs}.
This includes: virtual machines for both long/short studies and the
respective experimental/control groups~\cite{onlinevm}; source code
for the bugs chosen and corresponding patches; study instructions
for participants~\cite{onlineinstructions}, as well as tutorial
videos~\cite{onlinevideos}; and de-identified results (including
screen recordings)~\cite{onlinedata}.  These materials can be used
and modified to facilitate future experiment iterations.

%% file: figures/user_c_cpp_experience.tex
\begin{tabular}{llrll}
\toprule
 Study &     Treatment &  Subject ID &               C/C++ Experience \\
\midrule
  Long &       Control &        4 &         1 year  \\
  Long &       Control &        5 &         5 years \\
  Long &       Control &       11 &         2 years \\
  Long &       Control &       13 &         3 years \\
  Long &  Experimental &        2 &         1 year  \\
  Long &  Experimental &        3 &         5 years \\
  Long &  Experimental &       10 &         3 years \\
  Long &  Experimental &       12 &         2 years \\
 Short &       Control &        6 &         5 years \\
 Short &       Control &        7 &         2 years \\
 Short &  Experimental &        8 &         5 years \\
 Short &  Experimental &        9 &         3 years \\
\bottomrule
\end{tabular}

%% file: figures/expert_answers.tex
\begin{tabular}{lrr}
\toprule
    Treatment &  Subject &  \# Expert Answers \\
\midrule
      Control &        4 &                 4 \\
      Control &        5 &                32 \\
      Control &        6 &                35 \\
      Control &        7 &                33 \\
      Control &       11 &                 5 \\
      Control &       13 &                35 \\
 Experimental &        2 &                15 \\
 Experimental &        3 &                29 \\
 Experimental &        8 &                35 \\
 Experimental &        9 &                32 \\
 Experimental &       10 &                28 \\
 Experimental &       12 &                 6 \\
\bottomrule
\end{tabular}

%% file: findings.tex
\section{Findings}\label{sec:results}

Subjects in the experimental group did not display a significant
improvement relative to the control group. Both groups
submitted roughly the same number of correct patches. The
differences in time-to-first-submission across groups were
negligible.

\Cref{table:experimental:patches} describes the experimental group's
submissions relative to the available patches. Experimental subjects
in the long study submitted eight patches. Of these eight, four
were a direct use of a provided patch, and four were custom
patches. In the short study, subjects submitted four patches, of
which three were a direct use of a provided patch, and one was a
modification of a provided patch.

\begin{table}[]
\centering
\begin{tabular}{@{}llll@{}}
\toprule
Study & Direct Patch & Modified Patch & Custom \\ \midrule
Long  & 4                     & 0                     & 4               \\
Short & 3                     & 1                     & 0               \\ \bottomrule
\end{tabular}
\caption{%
Most experimental subjects provided a patch that was the direct
application of one of the provided patches. However, experimental subjects overwhelmingly picked the wrong patch. Only subjects in the
long study, where each bug was allocated 40 minutes, submitted custom patches.
}
\label{table:experimental:patches}
\end{table}

While most experimental subjects submitted one of the provided patches, few
chose the correct one. We inspected the screen recordings and characterize
behaviors that we believe drove these results
and present key challenges for
developer use of automatic patch generation systems.

\subsubsection{\bf Understanding Code Context}

Experimental subjects on average spent \fracinspectingvars{}
searching source code that contained variables
used in the automatically generated patches.  They searched for
declarations, definitions, and uses of variables across the entire
file and project, not simply in the area surrounding the defect
location.

\begin{figure}
% User 9, bug 8, [6:30]
\begin{subfigure}{\columnwidth} \includegraphics[width=\columnwidth,keepaspectratio]{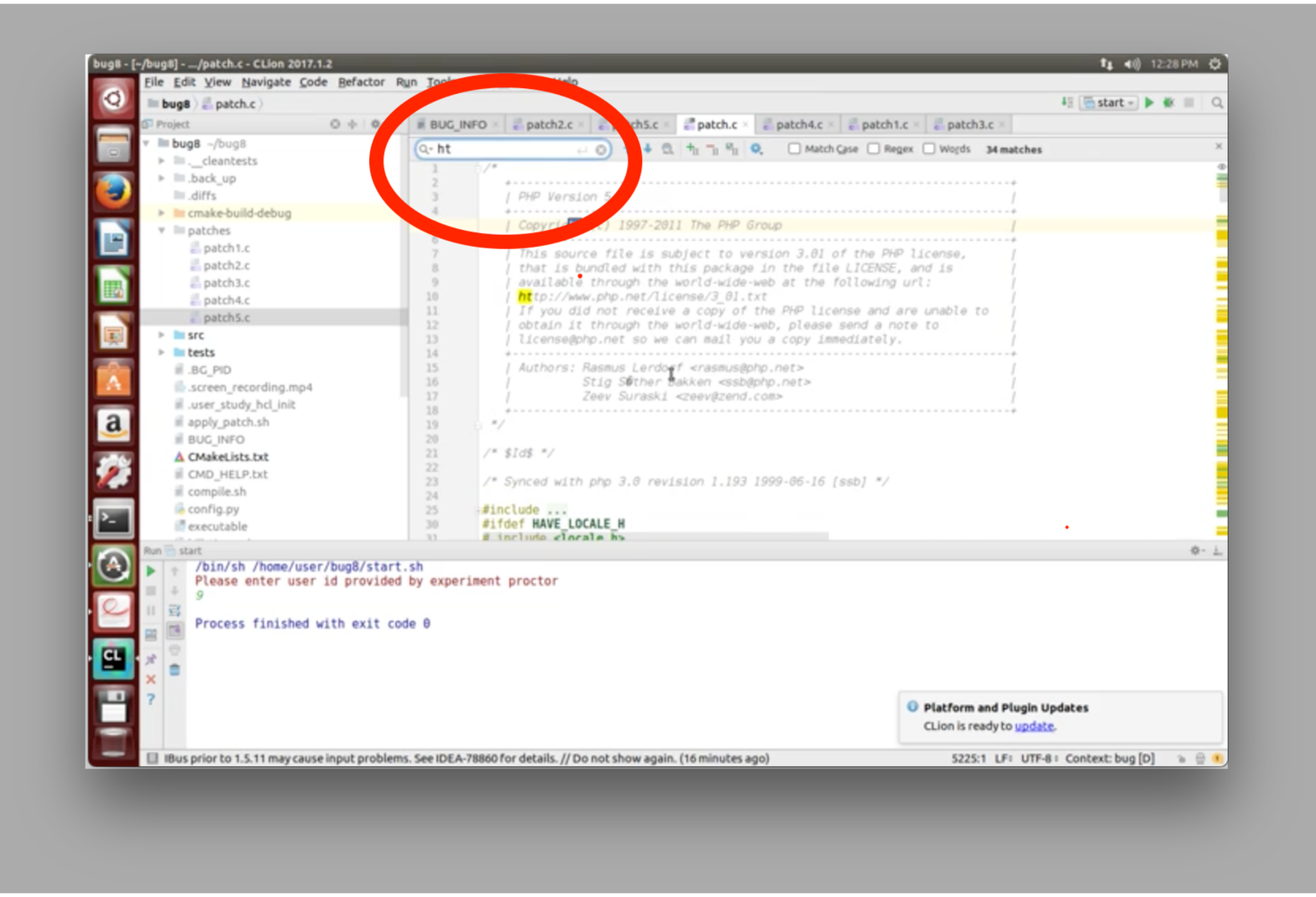}
\caption{%
Subject 9 searches for occurrences of variable \texttt{ht} in entire source file of \phpbug{}
}
\end{subfigure}
\begin{subfigure}{\columnwidth}
  \begin{lstlisting}[basicstyle=\footnotesize\ttfamily, breaklines=True]
if ((len > s1_len - offset) && !((ht == 4))) {
  len = s1_len - offset;
 }

cmp_len = (uint) (len ? len : MAX(s2_len, (s1_len - offset)));
  \end{lstlisting}
\caption{%
Patch provided for \phpbug{} that uses variable \texttt{ht}
}
\end{subfigure}
\caption{}
\label{fig:screenshot:variables}
\vspace{-1em}
\end{figure}

\Cref{fig:screenshot:variables} shows a subject searching
for instances of \texttt{ht} in the source file that contains the
bug. \texttt{ht} is one of the variables used in a patch
provided to the subject.

Experimental subjects spent \fracinspectingpatches{} inspecting the
provided patches. The remainder of the time was divided between
reading and modifying the original buggy source code, inspecting
tests, and reading the supplemental bug information.

We attribute both of these behaviors to experimental subjects
attempting to understand both the overall code they were trying to
repair and the interactions between the patch and the application.
This highlights the difficulty of patching application-logic bugs,
which require understanding the application semantics.

We expected subjects to leverage the patches more heavily, given that
the codebases for both bugs were unfamiliar to the subjects. The
relatively small fraction of time spent inspecting patches may indicate
that subjects prioritized understanding the application over picking
a repair that worked, but which they did not understand.

\subsubsection{\bf Patch Comparisons}
\begin{figure}
  % user 10, bug 2, [5:20]
  \includegraphics[width=\columnwidth]{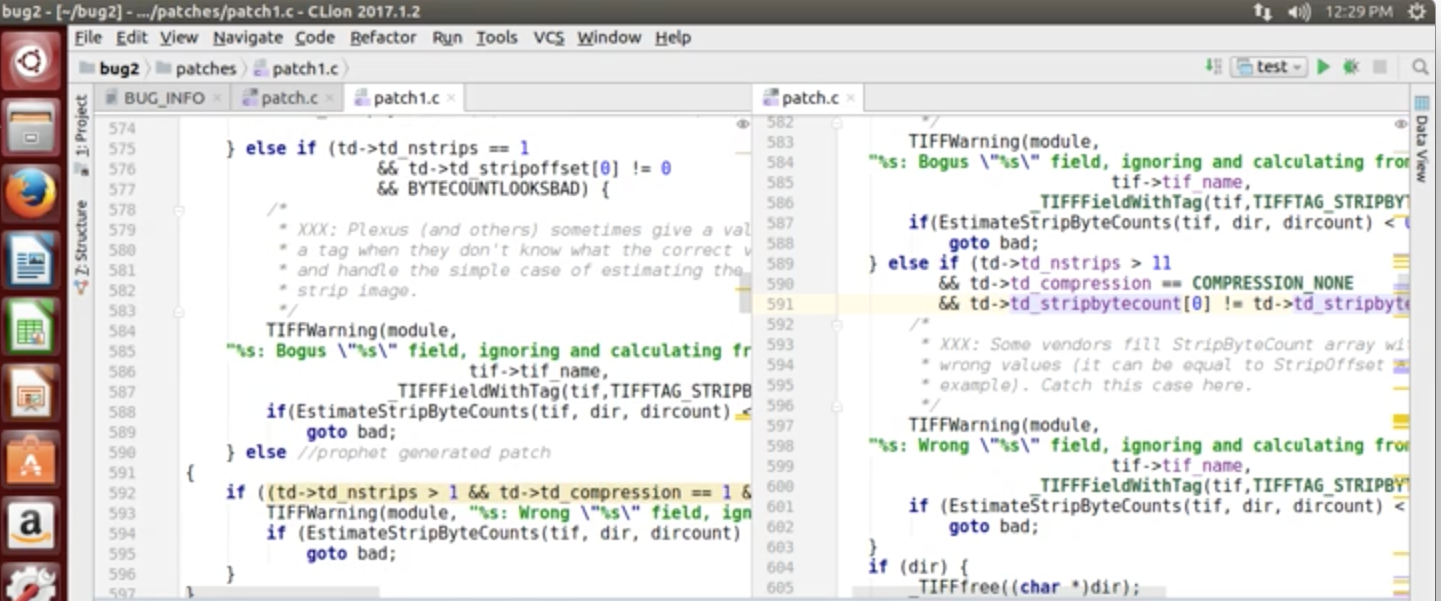}
  \caption{%
  Subjects used ad-hoc approaches to compare provided patches and
  the original source code. Subject 10, shown here, viewed provided patches side-by-side.
  Other subjects switched back and forth between files.
  }
  \label{fig:screenshot:comparisons}
\end{figure}

Reviewing videos of experimental subjects showed that the subjects used
a variety of ad-hoc approaches to compare different patches provided.
Some participants, such as subject 10 shown in \Cref{fig:screenshot:comparisons},
placed patches side-by-side and inspected the source code in both
files. Other subjects quickly navigated back and forth between files.

The lack of infrastructure dedicated to comparing differences across
patches may have complicated the code understanding task for subjects
as they explored patches. We observed subjects spent time switching
their screen use between multiple patches, and did so frequently
for patches that differed less between each other. The difference
between many of the patches provided to subjects was a single predicate.

\subsubsection{\bf Code Anchoring}

When comparing patches to the original buggy source code, we found
that subjects on average spent only \fracinspectingpatches{}
inspecting patches that we provided. Subject 9 spent the highest
fraction, \maxfracinspectingpatches{} of the allotted time on \libtiffbug{}.

Subjects spent less time inspecting patches that make changes that
may be considered significant.
Patch 1 in \\\libtiffbug{} and patch
3 in \phpbug{}, shown in \Cref{fig:patch:anchoring}, both add
\verb|false|
to the branch condition, which is equivalent to removing
the branch. For \phpbug{} this patch is the correct patch. We found
that on average experimental subjects only inspected these two patches
\fracdeletepatches{}.

\begin{figure}
% bug 8, patch3.c correct patch
\begin{lstlisting}[basicstyle=\ttfamily\footnotesize, breaklines=True]
if ((len > s1_len - offset) && !(1)) {
  len = s1_len - offset;
  }

cmp_len = (uint) (len ? len : MAX(s2_len, (s1_len - offset)));
\end{lstlisting}
\caption{%
Patches that made significant code changes to the original buggy
code were inspected less than patches
that made incremental changes. We found that patch 1 for
\libtiffbug{} and patch 3 for \phpbug{} (the correct patch), both of which
remove a branch, were only inspected for \fracdeletepatches{}.
}
\label{fig:patch:anchoring}
\vspace{-1.2em}
\end{figure}

We repeatedly informed experimental subjects
that the provided patches all passed the test
suites (i.e. validated), but that they were not guaranteed to be correct, and that
their task was to produce a correct repair. We also did not inform
the experimental group that there was a correct patch amongst the five
validated patches they received. However, in five of the 12
submissions from experimental subjects, subjects
did not apply additional patches nor make any changes after selecting
an initial patch that validated.  The single submission that modified
a validating patch did so by changing a comparison operation from
\lstinline{!=} to \lstinline{<} and submitted the new variant.
As a result of inspecting the patches for a relatively short period of time,
avoiding patches that seemed ``unlikely'', and not exploring the
application of multiple patches, we believe the experimental subjects
made relatively uninformed decisions for patch selection. This
highlights the challenges in selecting from several plausible patches.

%% file: implications.tex
%!TEX root = main.tex
\section{Implications For Developer-Centric Patch Generation Systems}
Our study illustrates that solely providing subjects with automatically generated patches may not be sufficient to see an effect in terms of patch integration productivity (measured by number of correct bug repairs or time to first patch).
Subjects spent most of their time trying to understand the defect and the way the provided patches related to the original source code containing the defect.

Based on our qualitative analysis, we formulate concrete directions for
future patch generation systems research. Particularly, providing mechanisms that help
developers better understand defects, and the relationship of the candidate
patch to the defects, can improve patch integration.\\
{\bf Variable Instrumentation.} Our qualitative analysis indicates that developers
    often spend time investigating the roles that variables from the generated patches play
    in the original defective code. Future systems could provide enhanced information about
    variables that occur in the generated patches, for example by providing program slices
    containing the code that affects the values of these variables~\cite{programslicing} or by
    providing dynamic information flow data that characterizes how these variables influence
    program outputs~\cite{sidiroglou2015automatic, sidiroglou2015targeted}.\\
{\bf Successful Patch Characteristics.} Developers in our study had difficulty
identifying correct patches in a set of plausible patches. Consequently, they
spent a lot of time trying to find contextual information to assess the
correctness of the given candidate patches. Machine learning has previously been
successfully used to identify characteristics of correct patches and
provide a probabilistic assessment of the viability of a
patch~\cite{long2016automatic}. Providing developers with this information can
aid patch integration as they inspect candidates, helping them to more
quickly distinguish correct from incorrect patches.\\
{\bf Trace and Influence Summaries.} Providing the developer with information about how
     the patch affects program execution characteristics, such as the flow of control
     and data through the program and output values, may make the potential impact
     of the patch clearer. This information would be collected during the runs of the original unpatched
     program and during the validation runs for candidate patches.\\
{\bf Invariants.}
Previous systems have inferred invariants that characterize successful executions~\cite{ernst2007daikon,kruger2006clearview}.
Providing invariants within the patch integration process that involve variables occurring in the patch may help developers better understand the roles that these variables play in the overall computation.

%% file: threats.tex
\section{Scope and Threats to Validity}
This study models what we believe are a crucial set of circumstances
that affect real-world use of a patch generation system. However,
the scope of our conclusions is limited by various design choices.

Our study tasked subjects with repairing two application-logic bugs
in unfamiliar codebases (which removed potential skill differentials
based on code familiarity). This setting naturally raises the difficulty of
patching bugs. A similar case of unfamiliar developers
is common in industry, where new or lateral hires are tasked
with working on a new codebase, or a new organization acquires
a contract to maintain an existing codebase for a fixed period
time. We believe our design models this
important use case for automatic patch generation systems, but
this also means our conclusions may not generalize to scenarios where
the subjects have deep codebase knowledge and may be better prepared
to identify correct patches.

Experimental subjects were not told that one of the patches was
correct. Instead, they were reminded that there were no
correctness guarantees. This may have influenced their behavior and
limited the amount of patches they were willing to explore, despite
our initial expectations that this would encourage subjects to try
more patches.

Subjects were not allowed to use additional debugging tools to control for
tooling experience. We chose bugs that we judged were reasonable to patch
without additional tools (e.g. no memory-related bugs). However, the lack of
access to external tools may have had an effect on the way subjects approached
the bugs and reasoned about possible fixes.

Control subjects received the exact defect location for bugs.
This models an ideal case of perfect error localization. We provided
the exact location of the bug to match the line modified by the
patches given to experimental subjects (and the ground truth developer patch)
in order to reduce confounding factors. Existing error
localization systems do not always produce the exact location of a
defect~\cite{pearson2017evaluating}, and so the generalizability of the control group
performance could vary under different error localization tools.

%% file: related.tex
\section{Related Work}
We present key related work in the area of automatic patch generation
research and developer-centric software systems.

{\bf Automatic Patch Generation. } A significant body of work in the software engineering community has
developed new techniques for automatically generating patches
~\cite{long2016automatic, mechtaev2016angelix, long2015staged, long2016genesis}.
However, a smaller subset of these studies
have included meaningful evaluation of developer interaction with such patches
through user studies. Our own work suggests that user studies can reveal
new directions for automatic patch generation research to improve usability.

Fry et al.~\cite{patchcomments} explored the relationship between
patch understandability and maintainability. In their study, subjects
answered code comprehension and maintenance questions. Their results found
that automatic patches were more maintainable. However, their study did not
consider the correctness of patches~\cite{qi2015analysis}.

Kim et al.~\cite{kim2013automatic} introduced PAR, an automatic patch generation
system as their main result. As part of the system evaluation, the authors
carried out a user study on the subjective ranking of patches based on
acceptability. Subjects were asked to rank patches produced by humans and by two
automatic patch generation systems: PAR and GenProg. They found a statistically
significant ranking difference between PAR and GenProg, and no such difference
for PAR patches with respect to the human-written patches. The authors did not
consider the correctness of the patches as part of the study.

Tao et al.~\cite{paruserstudy} present the only other study to evaluate the
impact of automatically generated patches on the number of correct
patches produced by developers. The study tasked subjects with
manually repairing five Java bugs in two hours.  Subjects in the
study were provided with either: one automatically generated patch
or method-level error localization information for the defect. The
patches provided could be of two quality levels, based on rankings
produced in prior work~\cite{kim2013automatic}. They found that automatically
generated patches of high quality improved the number of correct
patches produced by subjects.
Our study differs from the Tao et al. study in various key dimensions:

\begin{itemize}
  \item \textbf{Population Experts}:
  their study classified subjects as experts if they had programmed
  in Java for more years than the average subject (4.4 years).  All
  of our participants had at least four years of programming
  experience.  Our study asked subjects to answer a 42 question C concepts
  survey. We classified as experts subjects with more than 25 expert
  responses, and designed the study groups to balance the four
  non-expert subjects.

  \item \textbf{Single Patch Provided:} subjects in their study
  were given a single automatically generated patch. Our study
  provided experimental subjects with five plausible patches.

  \item \textbf{Patch Correctness:} the single high quality patch provided
  to experimental subjects in their study correctly repaired
  the defect in three of the five bugs. In our study subjects had
  four incorrect (but plausible) patches, along with a single correct
  patch, for each bug.

  \item \textbf{Error Localization:} control subjects in their study
  were told the method that contained the defect. Our study provided
  control subjects with the exact source line that contained the
  defect.  We provided exact defect location to compare the benefits
  of automatic patch generation over error localization.

  \item \textbf{Bug and Error Types:} subjects in their study
  were tasked with repairing five Java bugs: four Mozilla Rhino
  bugs and one Apache Commons Math bug. Three of the
  Rhino bugs produced \texttt{Null\-Pointer\-Exception}, one Rhino bug
  produced an \texttt{Array\-Index\-Out\-Of\-Bounds\-Exception}, and the Commons
  Math produced a custom \texttt{Convergence\-Exception}. Our study asked
  subjects to patch two C bugs from \verb|libtiff| and \verb|PHP|.
  Both bugs produced application-specific errors and required an
  understanding of the application semantics, rather than standard
  fixes.

\end{itemize}

%% file: hci_related_work.tex
{\bf Developer-Centric Software Systems. }
Research in the HCI community has developed frameworks to understand
software errors from the perspective of the developer~\cite{ko2005framework}
and investigate developer behavior and tool usage~\cite{myers2016programmers}.
Incorporating these insights into automatic patch repair can improve the
successful adoption of this technology.

Ko et al~\cite{ko2006exploratory} explore how developers debug in
unfamiliar applications, a task inherent in modern software development
where maintenance efforts can span much longer time frames than core
development. The analysis presented is closely aligned to our study,
which focused on unfamiliar applications and asked developers to produce a patch. Similar to our
study, the authors note that a significant amount of time was spent reading code,
and navigating between source files. Tooling that facilitates navigation of
unfamiliar code, searching for potentially relevant cues, and relates semantic
information such as dependences, could help developers during bug repair.

Suzuki et al~\cite{suzuki2017tracediff} present a system that allows
developers to quickly diagnose program errors by inspecting trace
differences between originally buggy and corrected code. This trace visualization
could be directly incorporated into automatic patch repair systems when presenting
potential repairs to developers. Trace differences, summarizing variable values
and incorporating value invariants, could help developers identify and eliminate
patches that validate but are incorrect.

Hoffswell et al~\cite{hoffswell2018augmenting} develop an alternative visualization
tool that incorporates graphical aids directly in the developers source code.
These aids can include small graphs that describe the value of variables
during program execution and at different snapshots. Our study revealed that
experiment participants spent a significant amount of time trying to understand
the role of variables in the application to be repaired. An ``situ-visualization''
of variable values could make this process more efficient.

Mikami et al~\cite{mikami2017micro} present a version control system for
small modifications often made in source code during exploratory programming
(``micro-versioning''). The system includes visual cues that enable the developer
to keep track of small changes. Program patches produced by generate-and-validate
systems, such as those used in our study, often contain small subtle differences.
A micro-versioning tool, such as that developed by Mikami, would allow
developers to efficiently compare patch differences and quickly revert/apply
changes during patch exploration. This stands in contrast to the side-by-side
source code comparisons that we observed subject participants using during our
study.

Starke et al~\cite{starke2009working} point out the fact that developers
will often inspect few code search results. Our study shows that developers
may behave similarly when inspecting automatically generated patches, spending
little time on each patch and less of that on patches that seem unlikely.
One of the implications for automatic patch generation system developers then
is to incorporate clear and effective patch ranking. Prophet~\cite{long2016automatic}
is a system that uses a probabilistic model for such a purpose.

%% file: conclusion.tex
\section{Conclusion}
Automated patch generation systems have been designed to solve the long-lasting
problem of software bugs.  However, humans remain an important component
in integrating the final patch to be applied. Thus a key area of research for
automatic patch generation is developer usability and productivity.
We provide an initial study to characterize the way developers use
automatically generated patches. Based on this study we formulate
possible research directions to improve developer adoption.

\section*{Acknowledgment}
The authors would like to thank all of the participants in the user
study for their time.